\documentclass{appolb}
\usepackage{graphicx}

\begin{document}
\title{Light-by-light scattering in ultraperipheral heavy-ion collisions
at the LHC%
\thanks{Presented at Excited QCD 2016}%
}
\author{Antoni Szczurek\footnote{Also at University of Rzesz\'ow, PL-35-959 Rzesz\'ow, Poland.}, Mariola K{\l}usek-Gawenda \\
and Piotr Lebiedowicz
\address{Institute of Nuclear Physics, Polish Academy of Sciences, Radzikowskiego 152,
PL-31-342 Krak\'ow, Poland}
\\
}
\maketitle
\begin{abstract}
We present cross sections for diphoton production
in (semi)exclusive $PbPb$ collisions, relevant for the LHC.
The calculation is based on equivalent photon approximation in
the impact parameter space.
The cross sections for elementary elastic scattering 
$\gamma \gamma \to \gamma \gamma$ 
subprocess are calculated including two mechanisms:
box diagrams with leptons and quarks in the loops and 
a mechanism based on vector-meson dominance (VDM-Regge) model
with virtual intermediate vector-like excitations of the photons.
We get measureable cross sections in $PbPb$ collisions. 
We present many interesting differential distributions which could
be measured by the ALICE, CMS or ATLAS Collaborations at the LHC.
We study whether a separation of box and VDM-Regge contributions
is possible. We find that the cross section for elastic
$\gamma \gamma$ scattering could be measured in the heavy-ion collisions 
for subprocess energies smaller than $W_{\gamma\gamma} \approx 15-20$~GeV.
\end{abstract}
  
\section{Introduction}

In classical Maxwell theory waves do not interact.
In contrast, in quantal theory photons can interact via quantal fluctuations.
So far only inelastic processes, i.e. production of hadrons or jets
via photon-photon fusion could be measured e.g. in $e^+ e^-$ collisions.

The light-by-light scattering to the leading and next-to-leading
order was discussed earlier in the literature, 
see \cite{Bohm:1994sf,Jikia:1993tc,Bern:2001dg}
also in the context of search for effects beyond Standard Model
\cite{Gounaris:1998qk,Gounaris:1999gh}.
The cross section for elastic $\gamma \gamma \to \gamma \gamma$ 
scattering is very small, so till recently it was beyond 
the experimental reach.
In $e^+ e^-$ collisions the energies and/or couplings of photons 
to electrons/positrons are rather small so that the corresponding 
$\gamma\gamma \to \gamma\gamma$ cross section is extremely small.
A proposal to study helicity dependent $\gamma\gamma \to \gamma\gamma$
scattering in the region of MeV energies with the help of high
power lasers was discussed recently e.g. in Ref.~\cite{Homma:2015fva}.

Ultraperipheral collisions (UPC) of heavy-ions provide a nice
possibility to study several two-photon induced processes such as:
$\gamma\gamma \to l^+l^-$, $\gamma\gamma \to \pi^+\pi^-$,
$\gamma\gamma \to $ dijets.
It was realized only recently that ultraperipheral heavy-ions 
collisions can be also a good place for testing elastic
$\gamma \gamma \to \gamma \gamma$ scattering experimentally 
\cite{d'Enterria:2013yra,KLS2016}.

In this communication we present our recent results obtained
in \cite{KLS2016}.
We shall show some differential distributions not discussed before
\cite{KLS2016}.
In Ref. \cite{KLS2016} we included both box mechanisms as well as a new soft 
mechanism relying on simultaneous fluctuation of both photons into virtual
vector mesons. This mechanism was not discussed before in the literature
in the context of photon elastic scattering. 

\section{$\gamma \gamma \to \gamma \gamma$ elementary cross section}

\begin{figure}[htb]
\centerline{%
\includegraphics[scale=0.3]{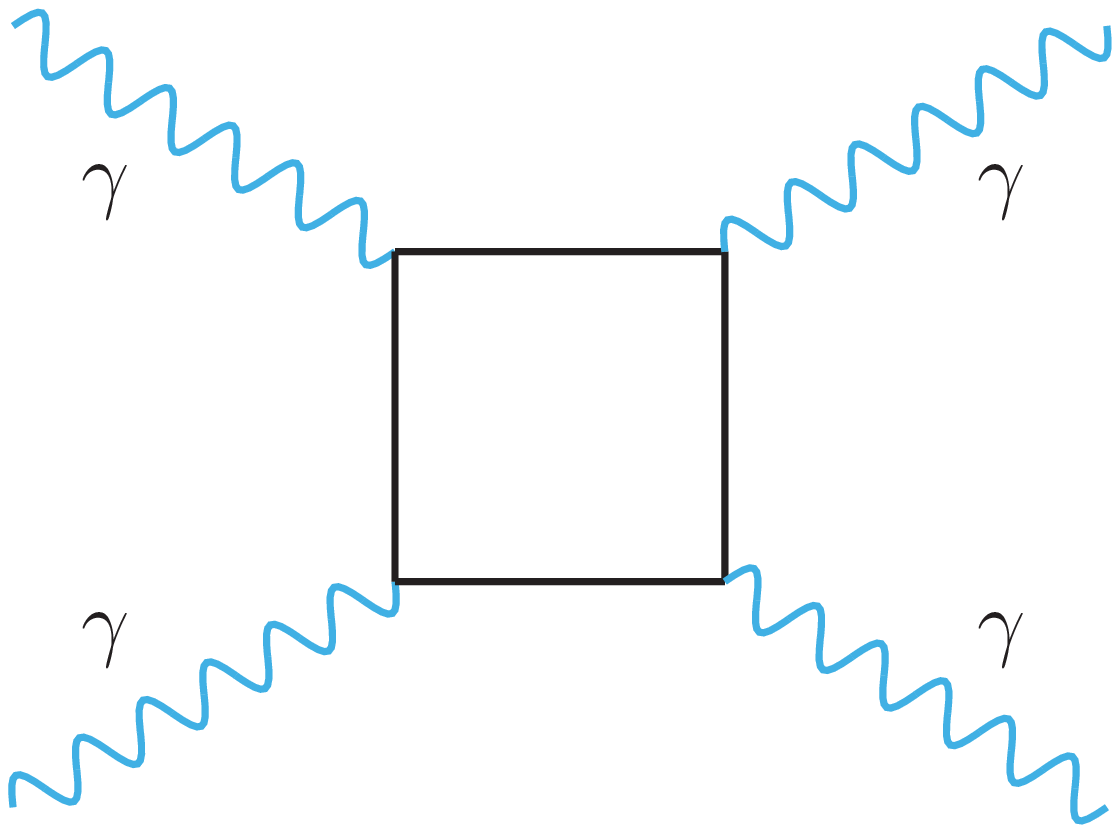}
\includegraphics[scale=0.3]{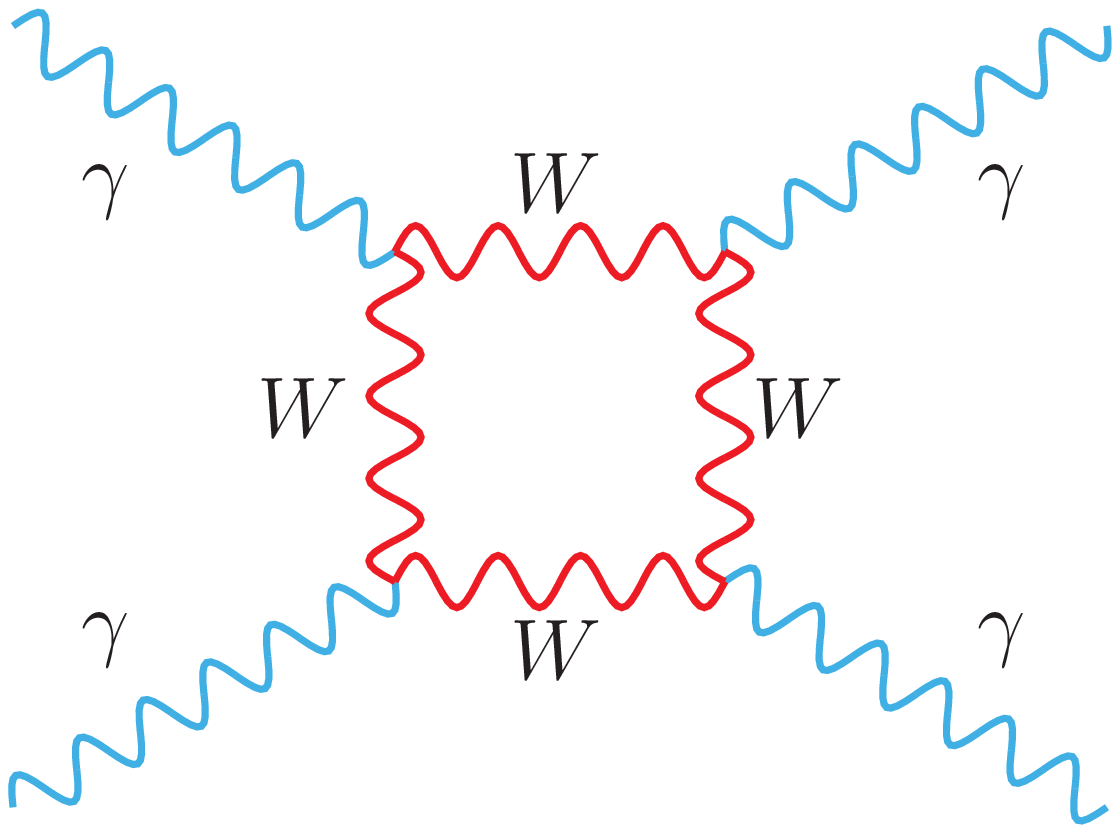}}
\caption{Light-by-light scattering mechanisms with 
the lepton and quark loops (left panel) and as an example 
one diagram for intermediate $W$-boson loop (right panel).}
\label{fig:diagrams_boxes}
\end{figure}

The lowest order QED mechanisms with elementary particles in the loop 
are shown in Fig.~\ref{fig:diagrams_boxes}. The diagram in the left panel
is for lepton and quark (elementary fermion) loops, while the diagram
in the right panel is for $W$ (spin-1) boson loops. 
The mechanism on the left hand side was shown to dominate
at lower photon-photon energies while the mechanism on the right hand 
side becomes dominant at higher photon-photon energies 
(see e.g.~\cite{Bardin:2009gq,Lebiedowicz:2013fta}).
In numerical calculations we include box diagrams with fermions
only.
The inclusion of $W$ bosons becomes important only at $W >$ 50 GeV
which, as will be shown below, is practically impossible
to observe at present with heavy-ion collisions.

\begin{figure}[htb]
\centerline{%
\includegraphics[scale=0.4]{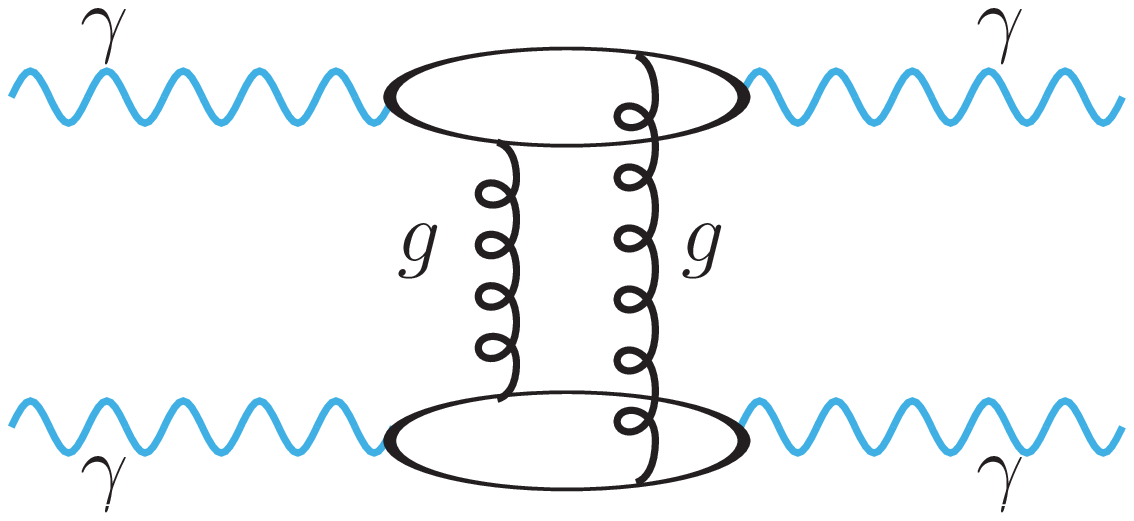}
\includegraphics[scale=0.4]{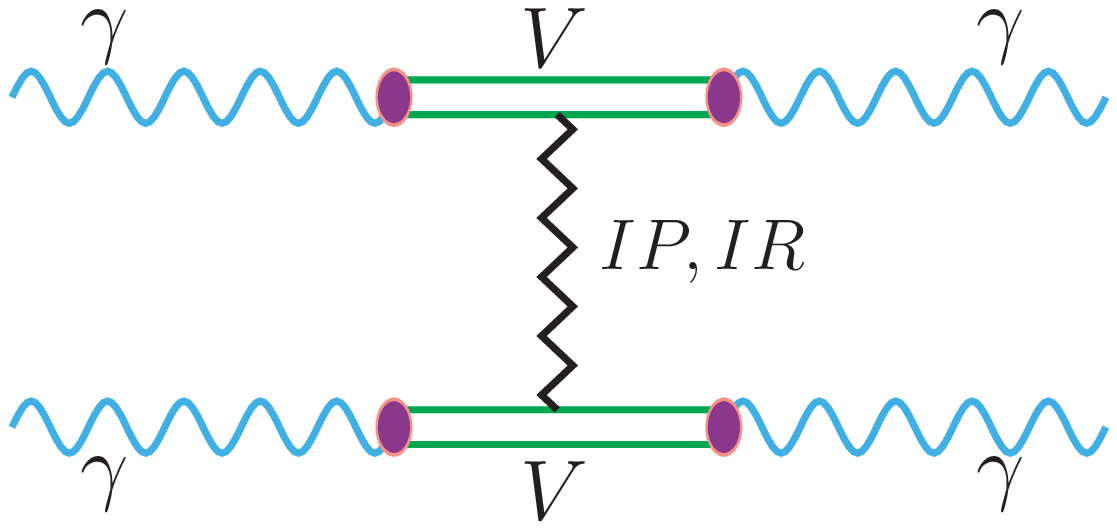}}
\caption{Other elementary $\gamma\gamma \to \gamma\gamma$ processes. 
The left panel represents two-gluon exchange and the right panel 
is for VDM-Regge mechanism.}
\label{fig:diagrams_t_channel}
\end{figure}

Two-loop corrections turned out to be small \cite{Bern:2001dg}.
However, higher-order processes are potentially interesting.
In Fig.~\ref{fig:diagrams_t_channel} (left panel) we show a process 
which is the same order in $\alpha_{em}$ but higher order in $\alpha_s$. 
This mechanism is formally three-loop type but can be calculated
in high-energy approximation \cite{KSS2016}.
Here we shall not discuss the higher-order contributions, instead we
shall discuss ''a kinematically similar'' process shown 
in the right panel where both photons fluctuate into virtual vector mesons
(three different light vector mesons are included). 
In this approach the interaction ''between photons'' 
happens when both photons are in their hadronic (vector-meson) states. 

Details of differential cross section and the amplitude 
for the VDM-Regge mechanism can be found in \cite{KLS2016}.

\begin{figure}[htb]
\centerline{%
\includegraphics[scale=0.275]{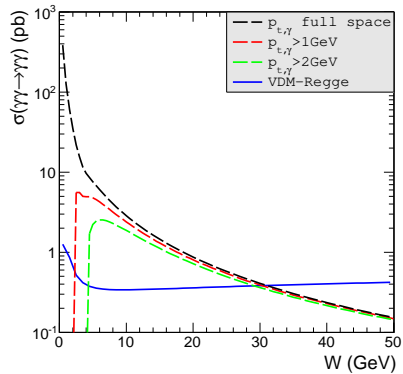}}
\caption{Integrated $\gamma\gamma \to \gamma\gamma $ cross section
as a function of the subsystem energy. 
The dashed lines show contribution of boxes and the solid line represents result
of the VDM-Regge mechanism.}
\label{fig:sig_elem}
\end{figure}

The elementary angle-integrated cross section for the box and 
VDM-Regge contributions is shown in Fig.~\ref{fig:sig_elem}
as a function of the photon-photon subsystem energy.
Lepton and quark amplitudes interfere in the cross section
for the box contribution.
At energies $W >30$~GeV the VDM-Regge contribution becomes larger
than that for the box diagrams.

\section{Production of pairs of photons 
in ultraperipheral heavy ion collisions}

At present (LHC) the photon-photon elastic scattering can be studied
in $p p \to p p \gamma \gamma$ and $A A \to A A \gamma \gamma$.
In Ref.\cite{KLS2016} we concentrated on heavy ion collisions.
Here we shall show some results obtained in \cite{KLS2016} for the
diphoton production in ultrarelativistic heavy ion collisions at the LHC.

\begin{figure}[htb]
\centerline{%
\includegraphics[scale=0.25]{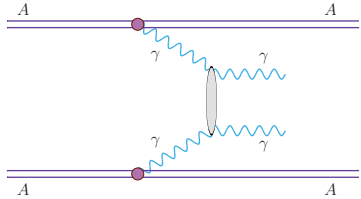}}
\caption{$AA \to AA \gamma \gamma$ in ultrarelativistic UPC of heavy
  ions.}
\label{fig:diagram_AA_AAgamgam}
\end{figure}
%
The general situation for $AA \to AA \gamma \gamma$ 
is sketched in Fig.~\ref{fig:diagram_AA_AAgamgam}.
In our equivalent photon approximation (EPA) in the impact parameter space, 
the total cross section 
is expressed through the five-fold integral
\begin{eqnarray}
\sigma_{A_1 A_2 \to A_1 A_2 \gamma \gamma}\left(\sqrt{s_{A_1A_2}}\right) 
&=&\int \sigma_{\gamma \gamma \to \gamma \gamma} 
\left(W_{\gamma\gamma} \right)
N\left(\omega_1, {\bf b_1} \right)
N\left(\omega_2, {\bf b_2} \right) 
S_{abs}^2\left({\bf b}\right)\nonumber \\
& \times & 
2\pi b \mathrm{d} b \, \mathrm{d}\overline{b}_x \, \mathrm{d}\overline{b}_y \, 
\frac{W_{\gamma\gamma}}{2}
\mathrm{d} W_{\gamma\gamma} \, \mathrm{d} Y_{\gamma \gamma} \;,
\label{eq:EPA_sigma_final_5int}
\end{eqnarray}
More details can be found in~\cite{KlusekGawenda:2010kx,KLS2016}.

\begin{figure}[htb]
\centerline{%
\includegraphics[scale=0.275]{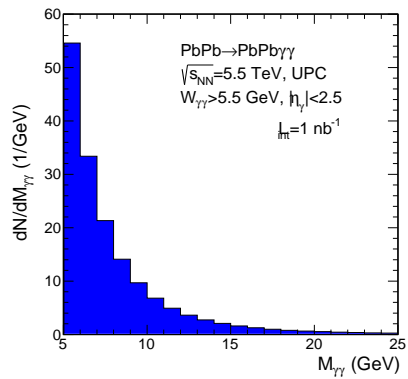}}
  \caption{\label{fig:number_of_counts}
  \small
Distribution of expected number of counts in $1$~GeV bins for cuts specified in the
figure legend.  This figure should be compared with a similar figure
in \cite{d'Enterria:2013yra}. 
}
\end{figure}

Can the elastic photon-photon collisions be measured with 
the help of LHC detectors?
In Fig.~\ref{fig:number_of_counts} we show numbers of counts
in the $1$ GeV intervals expected for assumed integrated luminosity 
of $1$~nb$^{-1}$,
where in addition to the lower cut on photon-photon energy
we have imposed cuts on (pseudo)rapidities of both photons.
It seems that one can measure invariant mass distribution up to 
$M_{\gamma \gamma} \approx 15$ GeV.

\begin{figure}[htb]
\centerline{%
\includegraphics[scale=0.275]{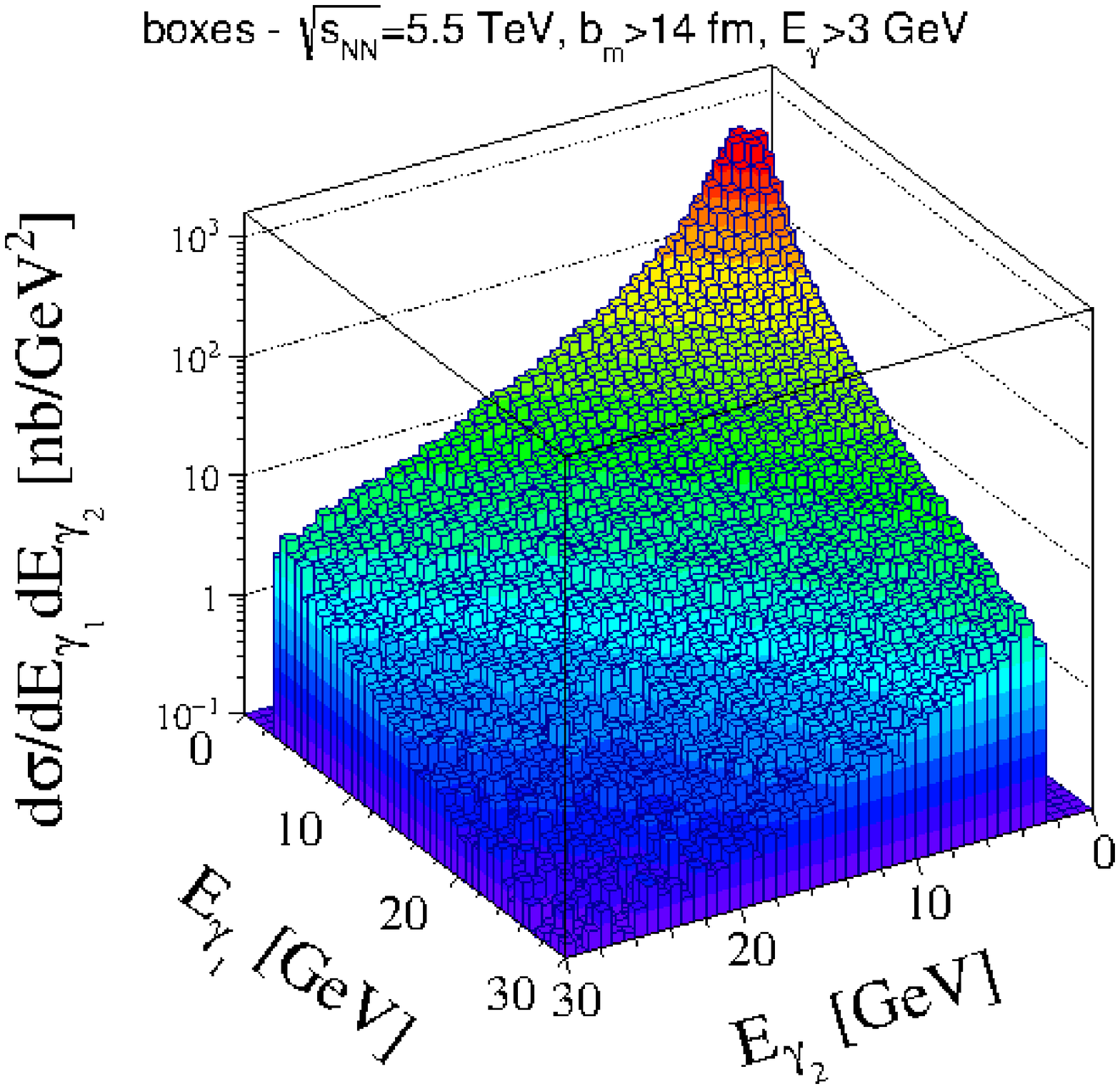}
\includegraphics[scale=0.275]{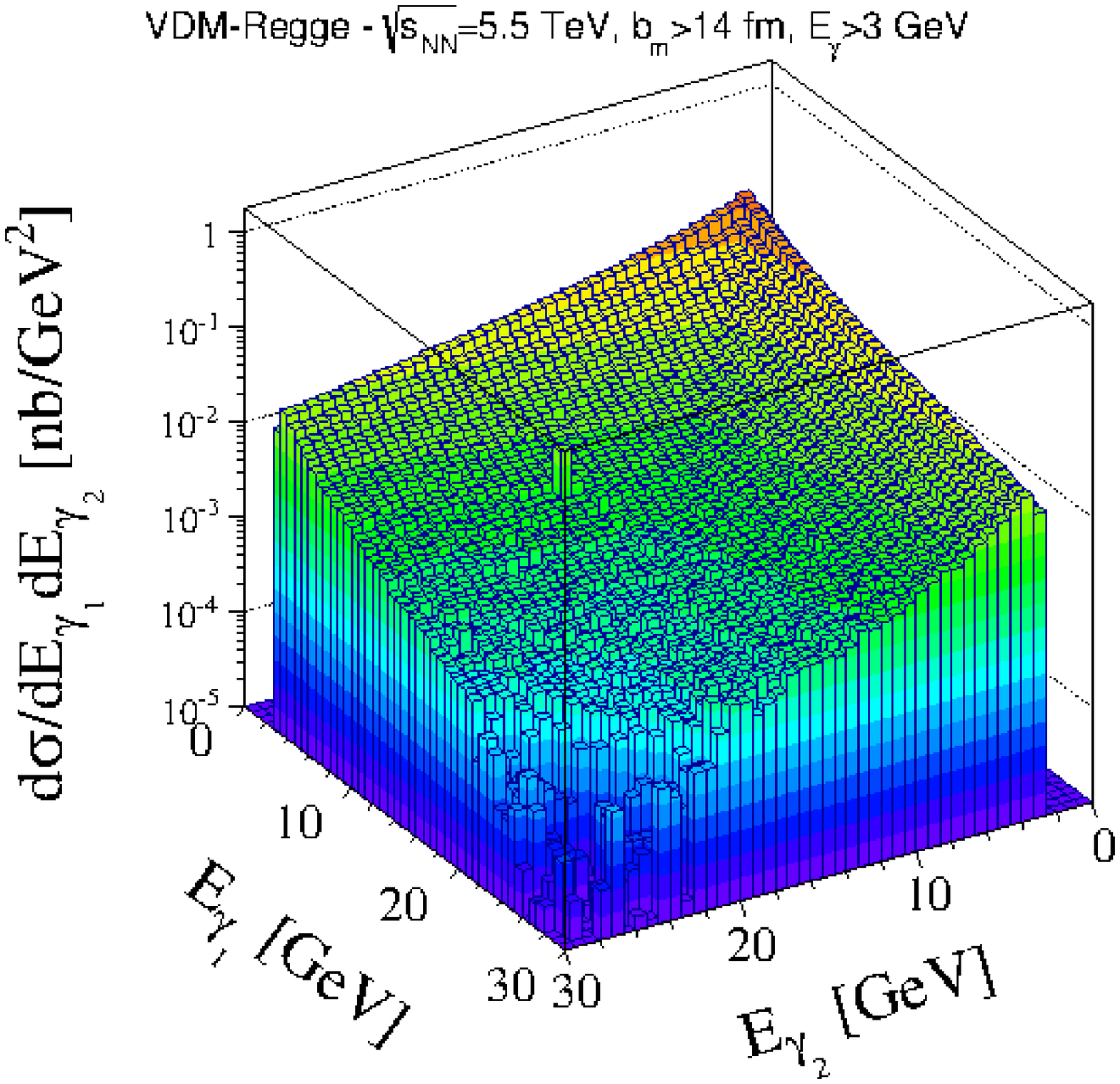}}
\caption{\label{fig:dsig_dE1dE2}
\small
Two-dimensional distribution in energies of the two photons in
the laboratory frame for box (left panel) and VDM-Regge (right panel) 
contributions.
}
\end{figure}

The cuts on subsystem energies are in principle not obligatory.
What are energies of photons in the laboratory frame?
In Fig.~\ref{fig:dsig_dE1dE2} we show distribution of energies
of both photons, separately for the two mechanisms: boxes (left panel)
and VDM-Regge (right panel). In this calculations we do not impose cuts
on $W_{\gamma \gamma}$ but only minimal cuts required by experiments 
on energies of individual photons ($E_{\gamma} > 3$ GeV).
Slightly different distributions are obtained for box and VDM-Regge
mechanisms. For the box mechanism we can observe a pronounced maximum
when both energies are small.
For both mechanisms one observes an enhancement of the cross section
for rather asymmetric configurations: $E_1 \gg E_2$ or $E_1 \ll E_2$.

\begin{figure}[htb]
\centerline{%
\includegraphics[scale=0.275]{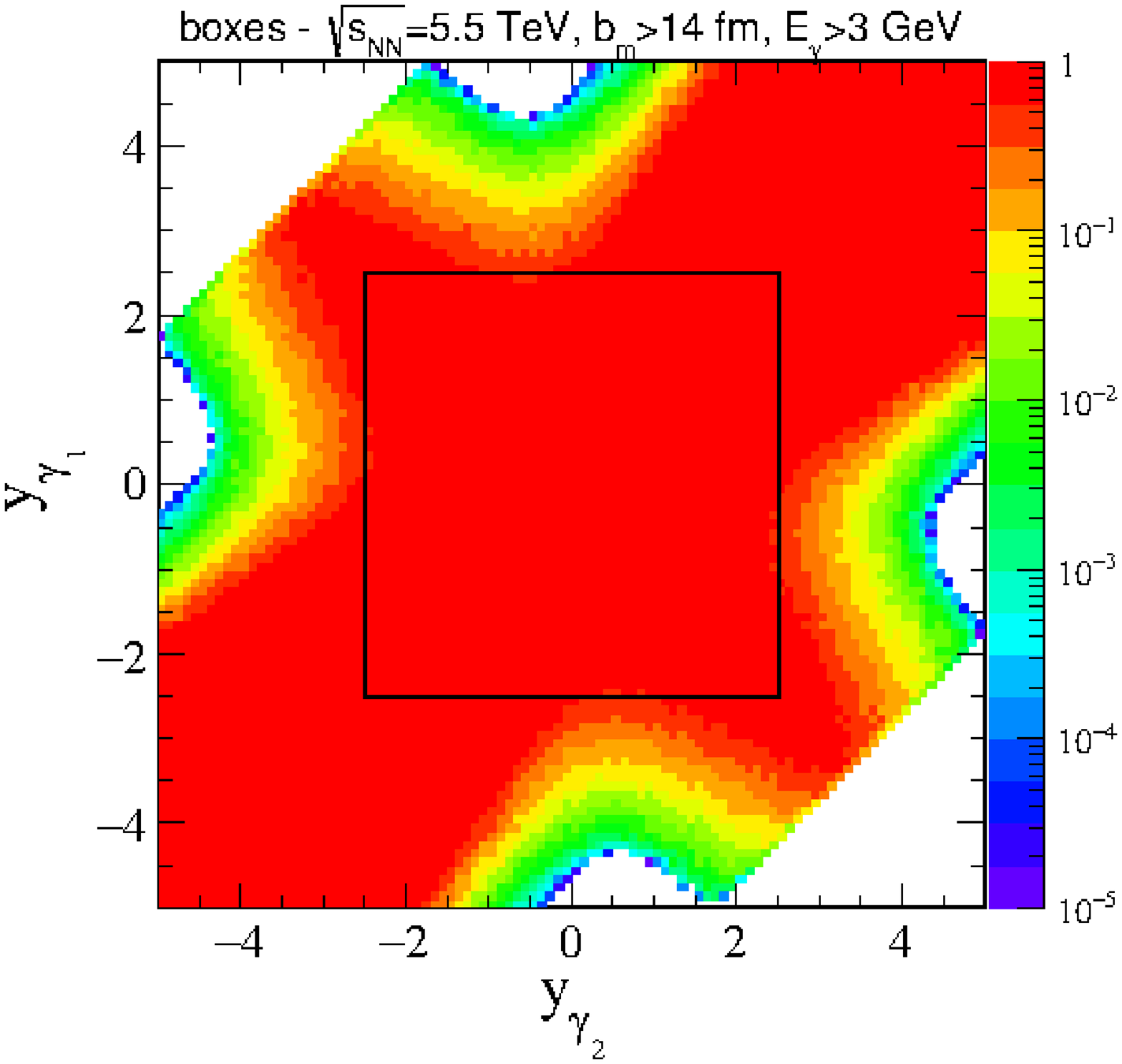}
\includegraphics[scale=0.275]{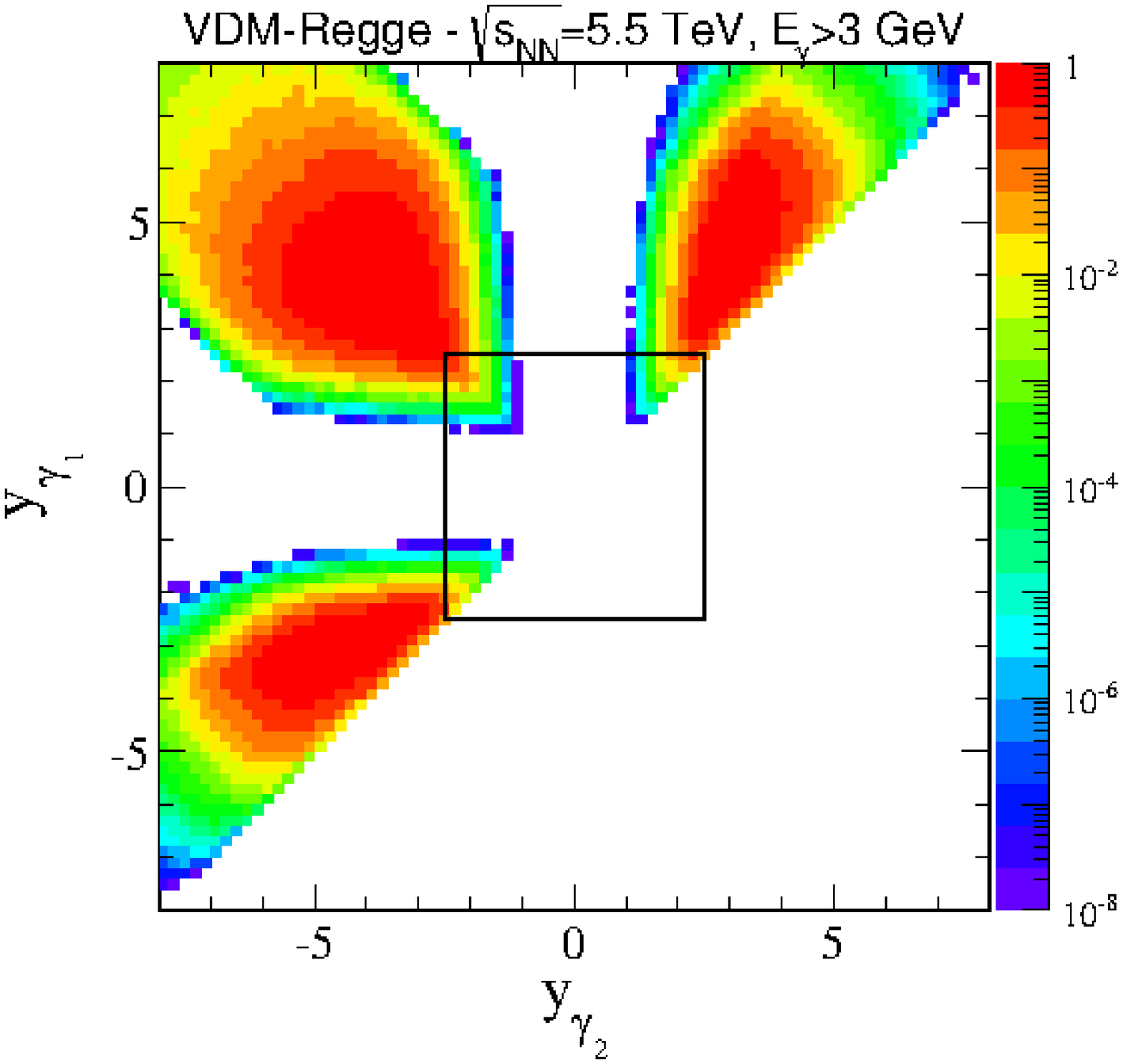}}
\caption{\label{fig:dsig_dy1dy2}
\small
Contour representation of two-dimensional 
($\mathrm{d} \sigma / \mathrm{d} y_{\gamma_1} \mathrm{d} y_{\gamma_2}$ in nb) 
distribution in rapidities 
of the two photons in the laboratory frame for box (left panel) 
and VDM-Regge (right panel) contributions together with experimental 
rapidity coverage of the main the ATLAS or CMS detectors.
Only one half of the ($y_{\gamma_1},y_{\gamma_2}$) space is shown for
the VDM-Regge contribution. The second half can be obtained by reflection
around the $y_{\gamma_1}=y_{\gamma_2}$ line.
}
\end{figure}

In Fig.~\ref{fig:dsig_dy1dy2} we show 
two-dimensional distributions in photon rapidities
with explicitly shown experimental limitations 
($y_{\gamma_1},y_{\gamma_2} \in (-2.5,2.5)$).
These distributions are very different for the box and VDM-Regge 
contributions. 
In both cases the influence of the imposed cuts is significant.
In the case of the VDM-Regge contribution we observe a non 
continues behaviour which is caused by the strong transverse momentum 
dependence of the elementary cross section
which causes that some regions in the two-dimensional space are 
almost not populated.
The empty areas in the upper-left and lower-right corners
for the box case are caused by a finite number of points in a grid at 
$z \approx \pm 1$.
For the case of the VDM-Regge contribution we show distribution for only
one half of the $(y_{\gamma_1},y_{\gamma_2})$ space. 
Clearly the VDM-Regge contribution does not fit
to the main detector and extends towards large rapidities.
In Ref.\cite{KLS2016} we investigated whether the photons originating
from this mechanism can be measured with the help of zero-degree 
calorimeters (ZDCs) associated with the ATLAS or CMS main detectors. 
In the case of the VDM-Regge contribution (right panel) we show much 
broader range of rapidity than for the box component (left panel). 
The maxima of the cross section associated with the VDM-Regge mechanism 
are at $|y_{\gamma_1}|,|y_{\gamma_2}| \approx$ 5. 
Unfortunately this is below the limitations 
of the ZDCs $|\eta| > 8.3$ for ATLAS (\cite{ATLAS:2007aa})
or $8.5$ for CMS (\cite{Grachov:2008qg}).

\section{Conclusion}

Recently in Ref. \cite{KLS2016} we performed detailed feasibility studies of 
elastic photon-photon scattering in ultraperipheral heavy ion collisions
at the LHC.
The calculation was performed in equivalent photon approximation
in the impact parameter space. This method allows to remove those
cases when nuclei collide and therefore break apart. Such cases
are difficult in interpretation and were omitted here.

In Ref.\cite{KLS2016} we proofed that the observation of the dominant
box contribution should be feasible as far as statistics is considered.
We also investigated whether the VDM-Regge contribution could be observed.
We observed that the VDM-Regge contribution reaches 
a maximum of the cross section when 
($y_{\gamma_1} \approx 5$, $y_{\gamma_2} \approx -5$)
or ($y_{\gamma_1} \approx -5$, $y_{\gamma_2} \approx 5$). 
This is a rather difficult region which cannot be studied e.g. 
with zero degree calorimeters installed at the LHC.

So far we have studied only two mechanisms of diphoton continuum.
The two-gluon exchange contribution, not discussed in Ref.
\cite{KLS2016} will be discussed soon in \cite{KSS2016}.
The resonance mechanism could be also included in the future.
In the present studies we have concentrated on the signal.
Future studies should include also estimation of the background.
The dominant background may be expected from 
the $A A \to A A e^+ e^-$ when both electrons are misidentified as
photons. 

\vspace{0.5cm}

This work was partially supported by the Polish grant 
No. DEC-2014/15/ B/ST2/02528 (OPUS).


\end{document}